\documentstyle[12pt,aasms4]{article}
\tighten

\newcommand\be{\begin{equation}}
\newcommand\ee{\end{equation}}

\input epsf
\lefthead{Basu et al.}
\righthead{Ring diagram analysis of near-surface flows in the Sun}
\begin{document}

\title{\bf Ring diagram analysis of near-surface flows in the Sun}

\author{Sarbani Basu}
\affil{Institute for Advanced Study, Olden Lane, Princeton, NJ 08540, U. S. A.} 
\author{H. M. Antia}
\affil{Tata Institute of Fundamental Research,
Homi Bhabha Road, Mumbai 400005, India} 
\and
\author{S. C. Tripathy}
\affil{Udaipur Solar Observatory, Physical Research Laboratory,
PO Box No. 198,\\ Udaipur 313 001, India}
\baselineskip=15pt

\begin{abstract}
Ring diagram analysis of solar oscillation power spectra obtained from
MDI data is carried out to study the velocity fields in the outer part of the
solar convection zone. The three dimensional power spectra are fitted
to a model which has a  Lorentzian profile in frequency and which includes
the advection of the wave front by horizontal flows, to obtain the two 
components of the sub-surface flows as a function of the horizontal wave
number and
radial order of the oscillation modes. This information is then inverted
using OLA and RLS methods to infer the variation in horizontal flow velocity
with depth. The average rotation velocity
at different latitudes obtained by this technique  agrees reasonably 
with helioseismic estimates made using frequency splitting data.
The shear layer just below the solar surface appears to consist of two
parts with the outer part up to a depth of 4 Mm, where the velocity gradient
does not show any reversal up to a latitude of $60^\circ$.
In the deeper part the velocity gradient shows reversal in sign around
a latitude of $55^\circ$.
The zonal flow velocities inferred in the outermost layers  appears to be
similar to those obtained by other measurements.
A meridional flow from equator polewards is found. It has a maximum amplitude 
of  about 30 m/s near the surface and the amplitude is nearly constant in
the outer shear layer.
\end{abstract}

\keywords{Sun: oscillations; Sun: rotation; Sun: interior}

\section{Introduction}

The rotation rate in the solar interior has been inferred using
the frequency splittings for p-modes (Thompson et al.~1996;
Schou et al.~1998). 
However, the splitting coefficients of the
global p-modes are sensitive only to the north-south axisymmetric
component of rotation rate. To study the non-axisymmetric component of
rotation rate and the meridional component of flow, other techniques
based on `local' modes are required.
Since these velocity components are comparatively small in magnitude
they have not been measured very reliably even at the solar surface.
The primary difficulty in measuring meridional flow velocities at solar
surface arises from convective blue shifts due to unresolved granular
flows (Hathaway 1987, 1992). Additional difficulty is caused by
the fact that at low latitudes the line of sight component of meridional
velocity is small. Sunspots and other magnetic features have also been
used to measure meridional flow (Howard 1996).
There is a considerable difference in the results of
these measurements. Using direct Doppler
measurements at the solar surface from GONG instruments Hathaway et al.~(1996)
have measured various components of nearly steady flows on the solar surface.
They find a polewards meridional flow with an amplitude of about 27 m/s, which
varies with time. There is also some evidence for north-south difference
in the rotation rate (Antonucci, Hoeksema \& Scherrer~1990; Verma 1993;
Carbonell, Oliver \& Ballester~1993; Hathaway et al.~1996)
but once again there is no agreement on the magnitude of this component or
its statistical significance.

Apart from these nearly steady flows, there could also be cellular flows with
very large length scales and life-times, viz., the giant cells.
However, there has been no firm
evidence for such cells (Snodgrass \& Howard 1984; Durney et al.~1985),
though recently Beck, Duvall \& Scherrer~(1998) have reported  probable
detection of
giant cells from the analysis of Michelson Doppler Imager (MDI) Dopplergrams.
These large scale flows are believed to play an important role in
transporting magnetic flux and angular momentum and thus, their study
is important
for understanding the theories of solar dynamo and turbulent compressible
convection (Choudhuri, Schussler \& Dikpati~1995; Brummell, Hurlburt \&
Toomre~1998; Rekowski \& R\"udiger 1998).

High-degree solar  modes ($\ell\ga150$)
which are trapped in the solar envelope have lifetimes that are much
smaller than the sound travel time around the Sun and hence the
characteristics of these modes are mainly determined by  average
conditions in local neighborhood rather than the  average conditions  over the
entire spherical shell. These modes
can  be employed to study large scale flows inside the Sun, using
time-distance analysis (Duvall et al.~1993, 1997; Giles et al.~1997),
ring diagrams (Hill 1988; Patr\'on et al.~1997) and other techniques.
Using  time-distance helioseismology Giles et al.~(1997) have studied
the meridional flow to find that the meridional velocity does not change
significantly with depth, while Schou \& Bogart~(1998) using the
ring diagram technique find some increase in meridional velocity with
depth.
Ring diagram analysis of meridional flows have also been done
by Gonz\'alez Hern\'andez et al.~(1998a) and Basu, Antia \& Tripathy~(1998),
who also find some variation in meridional flow with depth.
Haber et al.~(1998) find that the velocity of the surface flows
can change over moderately short time scales.
Gonz\'alez Hern\'andez et al.~(1998b) have demonstrated the reliability of
ring diagram analysis by comparing results obtained from data
collected simultaneously by two independent instruments, namely,
the MDI and one of the Taiwan Oscillation Network (TON) instrument
at Observatorio del Teide.

Ring diagram analysis is based on the study of three-dimensional
(henceforth 3d)  power spectra of solar
p-modes on a part of the solar surface. If one considers a section of a
3d spectrum at fixed temporal frequency, one finds that power is concentrated
along a series of rings that  correspond to different values of the
radial harmonic number $n$. The frequencies of these modes are 
affected by horizontal flow fields suitably averaged over the region under
consideration, hence, an accurate measurement of these frequencies
will contain the signature of large scale flows and can be used
to study these flows. The measured frequency shifts for different modes
can be inverted to obtain the horizontal flow velocities as a function of
depth. The local nature of these modes allows us to study different
regions on the solar surface, thus giving a three dimensional information
about the horizontal flows.
Since the high degree modes used in these studies
are trapped in the outermost layers of the Sun, such analysis gives
information about the conditions in the outer 2--3\% of the solar radius.

In this work we use  ring diagram analysis to study the longitudinal
as well as latitudinal component of horizontal velocity in the outer
layers of the Sun. Although, it is possible to study the variation in
horizontal velocity with both longitude and latitude, in this work
we have only considered the longitudinal averages, which contains information
about 
the latitudinal variation in these flows. For this purpose, at each latitude
we have summed the spectra obtained for different longitudes to get
an average spectrum which has information on the average flow velocity
at each latitude.
The longitudinal velocity component is dominated by the
rotation velocity and can be used to study its variation with depth and
latitude. This complements the results obtained from the inversion of
frequency splittings of global p-modes. Since the splittings of global
p-modes are not reliably determined at high degree these inversions are
not very reliable in regions close to surface.
It should, however, be  possible to determine the rotation rate in this region
more reliably with ring diagram analysis using data that include
high degree modes extending up to $\ell\approx1200$.
In particular, the shear layer
just below the solar surface can be examined in more detail.
Besides, ring diagram analysis also enables us to measure
the north-south variation in the rotation rate at same latitude.
The latitudinal component of velocity is dominated by the meridional
flow and can be used to study its variation with latitude and depth.
This work uses a  larger data set covering  an entire solar rotation period and
has improved fits as compared to the work reported in  Basu et al.~(1998).
We have also included an improved analysis of the meridional flow results
and identified higher order terms in this flow.

The rest of the paper is organized as follows: Section~2 describes the
basic technique used to calculate the horizontal flow velocities using
ring diagrams. Section~3 describes the results, while Section~4 gives
the conclusions from our study.

\section{The technique}

A good description of the ring diagram technique can be found in
Hill (1988, 1994) and Patr\'on et al.~(1997).  We, therefore,  only outline the
procedure used by us in this work. 
We have used data from full-disk Dopplergrams obtained
by the MDI instrument of the Solar Oscillations Investigation (SOI)
on board SoHO. Selected regions of
Dopplergrams mapped with Postel's projection are tracked at a rate
corresponding to the photospheric rotation rate (Snodgrass 1984) at the
center of each region to  filter out the
photospheric rotation velocity from the flow fields. This allows
us to study the smaller components of the flow which are not very
well determined from other studies.
Since the solar rotation rate generally increases with depth just
below the surface, the increase in rotation velocity will contribute
to the longitudinal component of horizontal flows measured by the ring
diagram technique.
For each tracked region, the images are
detrended by subtracting the running mean over 21 neighboring
images to filter the series temporally. Detrending eliminates slowly
varying signals such as those from local activity and nearly steady flows.
The detrended images are apodized and
Fourier transformed in the two spatial coordinates and in time to
obtain the 3d power spectra. We have chosen the spatial extent of
the region to be about $15^\circ\times15^\circ$ with $128\times128$ pixels
in heliographic longitude and latitude giving a resolution of
$0.03367$ Mm$^{-1}$ or 23.437 $R_\odot^{-1}$. Each region is tracked for
4096 minutes giving a frequency resolution of 4.07 $\mu$Hz.
To minimize effects of foreshortening all the regions were centered
on the central meridian, however, the high latitude regions will still
suffer from foreshortening.  Near the equator, each such region covers
an area of roughly $182\times182$ Mm on the solar surface and hence will
include a few supergranules. There
may still be some contribution from supergranular velocities in the
average flow over each region, which will interfere with the signal
from nearly steady and large  scale flows. Since in this work we are
considering only averages over all longitudes there will be further 
averaging of the  supergranular flows and their contribution is expected to
be very small.
The spectra have been obtained using the relevant tasks in MDI
data-processing pipeline.
Fig.~1 shows a few sections of some of these spectra at constant frequency.

We have selected the regions centered at Carrington longitudes of
$90^\circ, 60^\circ, 30^\circ$ for Carrington rotation 1909 and at
$360^\circ, 330^\circ, 300^\circ$, $275^\circ$, $235^\circ$, $210^\circ$,
$183^\circ$, $143^\circ$, and  $120^\circ$
for rotation 1910 corresponding
to a period from about May 24 to June 21, 1996, covering
an entire Carrington rotation period.
The uneven distribution of longitudes was dictated by the need to avoid,
as far as possible, large gaps in the data.
For each longitude
we select regions centered at latitudes of $60^\circ$ south
to $60^\circ$ north at  steps of $5^\circ$. Thus
there is some overlap between different regions.
Since in this work we are only interested in latitudinal variation in
the flow fields, we take a sum of all spectra for a given latitude
which gives us a spectrum averaged over all longitudes.
Because of averaging, these spectra
have better statistics and the error estimates are also lower.

To extract the flow velocities and other mode parameters from the
3d power spectra we fit a model of the form
\be
P(k_x,k_y,\nu)=
{\exp(A_0+(k-k_0)A_1+A_2({k_x\over k})^2+
A_3{k_xk_y\over k^2})\over
(\nu-ck^p-U_xk_x-U_yk_y)^2+(w_0+w_1(k-k_0))^2}+{e^{B_1}\over k^3}+
{e^{B_2}\over k^4}
\ee
where $k^2=k_x^2+k_y^2$, $k$ being the total wave number,
and the 12 parameters $A_0, A_1, A_2, A_3, c, p,
U_x, U_y, w_0, w_1, B_1$ and $B_2$ are determined by fitting the spectra
using a maximum likelihood approach (Anderson, Duvall \& Jefferies~1990).
Here $k_0$ is the central value of $k$ in the fitting interval.
The mean power in the ring is given by $\exp(A_0)/w_0^2$. The coefficient
$A_1$ accounts for the variation in power with $k$ in the fitting interval.
Only the linear term is included as the fitting interval is generally quite
small.
The $A_2$ and $A_3$ terms account for the variation of power along the
ring, namely, the variation with direction of propagation of the wave.
These were introduced because the power does appear to vary along
the ring and the fits in the absence of these terms were not satisfactory.
This variation can be easily seen in the outermost ring in the power spectra
displayed in Fig.~1. The outermost ring in the spectra around 3 mHz is
for $n=0$, while for spectra around 4 mHz the outermost ring is for $n=1$ as
the $n=0$ ring will be around $\ell=1600$, which is beyond the range of
our spectra.
The variation in power along the ring may be due to foreshortening or other
systematic effects and may not represent a real variation in the power
spectrum of the Sun.
The term $ck^p$ gives the mean frequency and this form is chosen as it
gives satisfactory fits to the mean frequency over the whole fitting
interval. The terms $U_xk_x$ and $U_yk_y$ represent the shift in
frequency due to large scale flows and the fitted values of $U_x$ and
$U_y$ give the average flow velocity over the region covered by the
power spectrum and the depth range where the corresponding mode is trapped.
The mean half-width is given by $w_0$, while $w_1$ takes care of the variation
in half-width with $k$ in the fitting interval.
The terms involving $B_1,B_2$ define the background
power, which is assumed to be of the same form as Patr\'on et al.~(1997).
The fitting formula given by Eq.~(1) is slightly different from what
is used by Patr\'on et al.~(1997), in that we have assumed some variation
in amplitude along the ring (given by  the $A_2$ and $A_3$ terms),
and we also include variation in power and width with
$k$ through the coefficients $A_1$ and $w_1$.
Basu et al.~(1998) did not include the $w_1$ term and the background terms
used by them were also slightly different.
Here, the positive $x$ direction is the direction of solar rotation and the
positive $y$ direction is towards the north in heliographic coordinates.

The fits are obtained by maximizing the likelihood
function $L$ or minimizing the function $F$
\be
F=-\ln L=\sum_i\left(\ln M_i+{O_i\over M_i}\right),
\ee
where summation is taken over each pixel in the fitting interval. The term
$M_i$ is the result of evaluating the model given by Eq.~(1) at
$i^{\rm th}$ pixel defined by $k_x, k_y, \nu$ in the 3d power spectrum,
and $O_i$ is the observed power at the same pixel.
The minimization has been performed using a
quasi-Newton method based on the BFGS (Broyden, Fletcher, Goldfarb and
Shanno) formula for updating the Hessian
matrix (Antia 1991). The error bars are obtained from the inverse of the
Hessian matrix at the minimum (Anderson et al.~1990).

In order to test the sensitivity of fits to the form of fitted function,
we have repeated the fits with some parameters kept fixed.
For example, we have tried fits keeping $w_1=0$ or $A_2=A_3=0$ and also
some fits where $w_0$ is kept fixed at the initial guess.
For background terms we have also tried exponents other than what are
included in Eq.~(1).
We have also attempted fits with more parameters including variation in
power due to higher order terms in $k_x, k_y$.
From these experiments we find that the fitted values of $U_x$ and $U_y$
are fairly robust to these changes and the differences between different
fits are less than the estimated errors. The form given by Eq.~(1) was
chosen because all the parameters appearing there have significant values
in some region and the fit appears to be satisfactory. While additional
parameters which were tried generally turned out to be small and 
comparable to the corresponding error estimates.
To evaluate the quality of the fit we use the merit function
(cf., Anderson et al.~1990)
\be
F_m=\sum_i\left(O_i-M_i\over M_i\right)^2
\ee
where the summation is over all pixels in the fitting interval and $O_i$
and $M_i$ are as defined in Eq.~(2). We find that with the choice of
model given by Eq.~(1), the merit function comes out to be close to unity
in all successful fits. If some of the parameters in Eq.~(1) are kept fixed
then the merit function increases. On the other hand, adding more parameters
does not reduce the merit function significantly.

We fit each ring separately by using the portion of power spectrum
extending halfway to the adjoining rings. For each fit a region extending
about $\pm100\mu$Hz from the chosen central frequency is used.
We choose the central frequency for fit in the range of 2--5 mHz
as the power outside this range  is not significant. The
rings corresponding to $0\le n\le6$ have been fitted.
For each value of $n$ we increase the central frequency of the fitting
interval in steps of 12.21 $\mu$Hz or 3 pixels in the spectra.
This gives us typically 800 `modes',   all of which  may not be independent
as there is a considerable overlap between adjacent fitting intervals.
In this work we express $k$ in units of $R_\odot^{-1}$, which enables
us to identify it with the degree $\ell$ of the spherical harmonic of
the corresponding global mode. 

The fitted $U_x$ and $U_y$ for each mode represents an average
--- over the entire region in horizontal extent and over the vertical
region where the mode is trapped --- of the velocities in the $x$
and $y$ directions respectively.
We can invert the fitted $U_x$ (or $U_y$) to infer the variation in
horizontal flow velocity $u_x$ (or $u_y$) with depth.
We use the Regularized Least
Squares (RLS) as well as  the Optimally Localized Averages (OLA)
(Backus \& Gilbert 1968) techniques for inversion. The results obtained
by these two independent inversion techniques are compared to test the
reliability of inversion results.

In the  RLS method we try to fit 
$U_x$ (or  $U_y$) under the constraint that the underlying $u_x$
(or $u_y$) is smooth. We represent $u_x$ (or $u_y$) in terms of 
a cubic B-Spline basis and the coefficients are determined by 
$\chi^2$ minimization with first derivative regularization.

In the OLA  technique the aim is to explicitly form linear combinations
of the data and the corresponding kernels such that the resulting
averaging kernels
are as far as possible localized near the position for which the solution
is being sought. This is done by
minimizing
\be
\int (r-r_0)^2 \left(\sum_i c_i K_i\right)^2 dr + \mu\sum_i c^2_i\sigma^2_i,
\ee
where $K_i$ are the mode kernels, $\mu$ is a trade-off parameter that 
ensures that the propagated
errors in the solution are low. The minimization is done subject to
the condition that the averaging kernel,  defined as
${\cal K}(r)=\sum_i c_i K_i(r)$, is unimodular, i.e., $\int {\cal K}(r) dr=1$.

For the purpose of inversion, the fitted
values of $U_x$ and $U_y$ are interpolated to the nearest integral
value of $k$ (in units of $R_\odot^{-1}$) and then the kernels
computed from a full solar model with corresponding value of degree
$\ell$ are used for inversion. Since the fitted modes are trapped in
outer region of the Sun, inversions are carried out for $r>0.97R_\odot$
only.

\section{Results}

Following the procedure outlined in Section~2 we fit the form given by Eq.~(1)
to a suitable region of a  3d spectrum. Fig.~2 shows some of the fitted
quantities for the averaged spectrum centered at the equator.
The error bars are not shown for clarity.
The power is maximum around a frequency of 3 mHz for modes with $n>1$.
The fitted half-width $w_0$ appears to increase at low frequencies. This
increase is probably artificial and we have checked that keeping the width
fixed during the fit for these modes does not affect the fitted $U_x$
and $U_y$.
Although not shown, we find  that the parameters $A_1, A_2, A_3$ defining the variation
in power with $k,k_x,k_y$  all have significant values.
The parameters $A_2$ and $A_3$
increase significantly with $k$ and this can be seen from the power spectra
shown in Fig.~1, where the variation of power along the ring is clearly
visible in the outer rings. 
In general only one of the two background terms defined by $B_1$ and $B_2$ is
significant, with $B_2$ being the more dominant at higher $\ell$.
 It thus appears that the
background decreases more rapidly with $\ell$ at higher $\ell$.
In principle, the
mean frequency $\nu_0=ck^p$ can also be computed from the fits, but there
would be some systematic errors in these values, as in other ridge fitting
techniques (Bachmann et al.~1995). The exponent $p$ varies between 0.35 to
0.55 for various modes. For f-modes it is in general very close to 0.5 which
is the expected asymptotic value for a standard solar model.

Although other quantities may also be of some interest, in this work we
restrict our attention to the two horizontal components of velocity obtained
by fitting the spectra. These are shown in Fig.~3 for various latitudes.
From this figure it appears that $U_x$ generally increases with depth, except
possibly at high latitudes. On an average, $U_x$ appears to be
lower at high latitudes, mainly due to the $\cos(\theta)$ (where
$\theta$ is the latitude) factor in conversion
from angular velocity to linear velocity. The latitudinal component $U_y$
is positive in the northern hemisphere and negative in the southern hemisphere
and thus the meridional flow is directed from equator to poles. Further,
the meridional component
appears to be comparatively independent of depth at low latitudes, while at
high latitudes there is some variation with depth.
The fitted velocities for each `mode' are inverted to obtain the
variation of horizontal velocity with depth. Only the region $r>0.96R_\odot$
is sampled by the modes used in this study and hence the inversions are
restricted to $r>0.97R_\odot$ as below this depth the averaging kernels are
not properly localized.
A sample of the averaging kernels for OLA inversion are shown in Fig.~4. 
Note that by $r=0.97R_\odot$,
the averaging kernels become wide and thus have poor resolution.
Also note that the peak of the averaging kernel for
0.9987$R_\odot$ is shifted slightly inwards, this is because there are
very few modes in the data set with turning points in that region.

\subsection{The rotation velocity}

From the inversion results it appears that the longitudinal component
($u_x$) is dominated by the average rotational velocity.
This is due to the fact that tracking is done at the surface rotation
rate at the center of the tracked region and hence does not account for the
variation of the rotation rate with depth. This velocity can be
compared with the helioseismic estimate after subtracting out the surface
velocity used in tracking. 
We find that there is a reasonable agreement between
$u_x$ and the rotation rate determined from splitting coefficients of the
global p-modes. Thus, this provides a test of our procedure for inferring the
subsurface velocity components. The results obtained using the RLS and OLA
techniques for inversion also agree with each other to within the
estimated errors.

The rotation velocity at each latitude can be
decomposed into the symmetric part [$(u_N+u_S)/2$] and an
antisymmetric part [$(u_N-u_S)/2$]. The symmetric part can be compared
with the rotation velocity as inferred from the splittings of global modes
(Basu \& Antia~1998) which sample just the 
symmetric part of the flow.  The comparison is shown in Fig.~5.
Since the inversion results using global modes which are restricted to a
mode-set of $\ell\le 250$ are not
particularly reliable in the surface regions, the velocity profiles obtained
from the ring diagram analysis supplement those results and support the
earlier conclusions. 

The shear layer just below the surface where the rotation rate increases
with depth has been seen in inversion results from global p-modes
(Schou et al.~1998; Antia, Basu \& Chitre 1998).
Some of these inversion results show a change of sign in the velocity
gradient just below the solar surface at high latitudes. 
It would be interesting to study this shear layer using ring diagram analysis.
From the results shown in Fig.~5,
it appears that the there is some tendency for reversal of gradient near
the surface at high latitudes.
This tendency for reversal is probably not significant since
this feature is seen in the region $r\ga 0.998R_\odot$,
and at these high latitudes
there are very few modes with lower turning
points in this region (cf., Fig.~3) making the inversion results less
reliable. To some degree this feature is seen at all latitudes. At lower
latitudes the extent is somewhat less, which supports the view that this
is due to lack of modes with lower turning point in the outermost region.
Spectra from high resolution Dopplergrams will help in the study of
this region.  Leaving aside this region,
it appears that the shear layer actually consists of
two layers, in the outer layer ($r\ga 0.994R_\odot$ or depth $\la 4$ Mm) the
gradient is in general quite steep and continues till the latitude
of $60^\circ$.
While in the second layer below this depth the gradient is smaller and
appears to reverse its sign around $55^\circ$ latitude.
The results obtained from global p-modes from MDI data shows a reversal
in the sign of the gradient
at latitudes of $55^\circ$. It is possible that this is due to relatively
poor resolution of global p-modes in this region. It is quite likely,
that the splittings of p-modes do not resolve the outer part of shear layer
and only the change in gradient in the deeper layer is extrapolated to
the surface.

The difference between rotation velocity at the
same latitude in the north and south hemispheres is small and thus
the antisymmetric component of rotation rate may not be very significant.
Some of the
difference may also be due to some systematic errors in our analysis.
For example, due to differences in the angle of
inclination for the regions at same latitude in the two hemisphere, the
effect of foreshortening will be different in the two hemispheres.
The antisymmetric component of the rotation velocity is shown in
Fig.~6. In particular, it can be seen that at low latitudes where the
results are more reliable, this component is generally small,
being comparable to the error estimates. Fig.~7 shows the antisymmetric
component plotted
as a function of latitude at two different depths.
This can be compared with the inferred value at the surface
from GONG data (Hathaway et al.~1996) as shown by Kosovichev \& Schou (1997).
Near the surface,
this component appears to be significant around latitude of 20--30$^\circ$.
In deeper layers the errors are larger and it is difficult to judge
the significance of this component.

As has been done by  Kosovichev \& Schou (1997), it is possible to
decompose the rotation velocity into two components, a smooth part
(polynomial in terms of $\cos(\theta)$, $\cos^3(\theta)$ and
$\cos^5(\theta)$, $\theta$ being the latitude)
and the residual which has been identified with zonal flows.
It may be noted that there is some ambiguity here since unlike the flow
found by Kosovichev \& Schou, the rotation velocity inferred from ring
diagrams also includes the antisymmetric component. Thus it is not clear
if the antisymmetric terms also need to be included in the smooth part.
However, since the rotation velocity is traditionally
expressed using these three terms we have used this form for the smooth
component, though  this may result in the addition of the  antisymmetric
component to the zonal
flow pattern.  The zonal flow so estimated is shown in Fig.~8. 
The inferred pattern  near the surface
is similar to the average zonal flow estimated from the
splitting coefficients for the f-modes from the 360 day MDI data.
The agreement appears to be better in the northern latitudes.
This pattern can also be compared with Fig.~3 of Hathaway et al.~(1996),
which shows the zonal flow at the solar surface as inferred from Doppler
measurements. This includes the antisymmetric component also and hence
it is more meaningful to compare our results with the GONG measurement
at solar surface. There is good agreement between the OLA and RLS results.
It may be noted that the error bars for zonal flows shown in Fig.~8 
are just the errors in determining the rotation velocity at each latitude.
Additional errors may arise from uncertainty in the smooth component which
is subtracted out to obtain these values. Thus the errors may have been
underestimated.
One must keep in mind
that the global f-modes are only sensitive to the north-south
symmetric component of zonal flows and if average is taken for the
ring diagram results over the north and south latitudes,
then the agreement is better (Fig.~9).
There is also some variation with depth in the zonal flow pattern
as can be seen from Fig.~8, while the f-mode results represent some
average over the region at depths of 2--9 Mm (Kosovichev \& Schou 1997).
At deeper depths the pattern changes and the errors are
also larger. Hence, it is not clear if
the zonal flow penetrates below about 7 Mm ($0.01R_\odot$) from the surface.

\subsection{The meridional flow}

The latitudinal component of the velocity appears to be dominated by the
meridional flow from equator polewards. The average latitudinal velocity
for each latitude is shown in Fig.~10, while Fig.~11 shows the same
as a function of latitude at a few selected depths.
There is a significant variation in this velocity with depth at high
latitudes.  Since the measurements may not be very
reliable at high latitudes it is difficult to say much about the
general form of the flow velocity with latitude by looking at these figures.
In any case, it
also depends on depth. Thus in order to understand the variation of
meridional flow velocity with depth and latitude we attempt to fit 
a form (cf., Hathaway et al.~1996)
\be
u_y(r,\phi)=-\sum_i a_i(r) P_i^1(\cos(\phi))
\ee
where $\phi$ is the colatitude, and $P_i^1(x)$ are the associated Legendre
polynomials.
The odd terms in this expansion give the north-south symmetric
component, while even terms give the dominant anti-symmetric component.
The variation of amplitudes $a_i(r)$ with depth will give the depth
dependence of the flow velocity.
The first two terms in this expansion are $cos(\theta)$, and
$(3/2)\sin(2\theta)$, where $\theta$ is the latitude.
These are same as those used by Giles et al.~(1997) to fit the meridional
flow velocity obtained from time-distance analysis. We find that these
terms are not sufficient and it is necessary to include about 6 terms
before the fits look reasonable. The terms beyond the sixth are found to
be smaller than the respective error estimates at all depths.
The amplitudes of the first six terms are shown in Fig.~12.

The second term is  the largest with an amplitude (after
accounting for the factor of 3/2) of 20--35 m/s depending on the
depth. The odd components representing the symmetric part are generally small
being comparable to error estimates, except in the region $r>0.99R_\odot$
where the values appear to be somewhat significant.
It is not clear if a part of these terms is due to some systematic errors
arising from misalignment in the MDI instrument (Giles et al.~1997).
The coefficient $a_4$ appears to be significant in the intermediate depths
with maximum value of 3--4 m/s. This component has been suggested by
Durney~(1993) from theoretical considerations involving differential
rotation. While Giles et al.~(1997) did not find significant value for this
component, we find that although near the surface $a_4$ is small it becomes
significant in deeper layers. The coefficient $a_6$ is even smaller,
though the value may be significant at some depths.

The amplitude of the dominant component ($\sin(2\theta)$)
is about 30 m/s near the surface,
which is comparable to the values obtained by Giles et al.~(1997)
from time-distance analysis, and by Hathaway et al.~(1996) from direct
Doppler measurement at solar surface. The form of meridional velocity fitted
by Hathaway et al.\ is same as what we have used, but they have not published 
the higher components, which are probably small at the surface.
On the other hand, Giles
et al.\ have used only the first two components and consequently it is
not clear if the results can be directly compared since the higher order
polynomials  also have terms of form $\sin(2\theta)$, which also make
some contribution to the amplitude.

We find the amplitude ($a_2$) of the dominant component of meridional flow
velocity is roughly constant near the surface, but decreases around
$r=0.994R_\odot$, below $r=0.99R_\odot$ the amplitude again increases.
It may be noted that the region $r>0.994R_\odot$
where the amplitude is nearly constant coincides with the region where we
identified the outer shear layer in rotation velocity. The other coefficients
$a_4$ and $a_6$ of meridional flow also show a change in amplitude around
this depth. This reinforces our conclusion about two different shear
layers below the solar surface.
Near the surface and again at depths of around 21 Mm, the coefficients
$a_4$ and $a_6$ are small and the meridional flow profile shows a decrease
at higher latitudes as expected from the second term. At intermediate depths
there is no sign of this turnover in velocity up to latitudes of $60^\circ$.

There is no evidence for any change
in sign of the meridional velocity up to a depth of $0.03R_\odot$ or
21 Mm that is covered in this study. Thus the return flow from the poles
to equator must be located at greater depths.

\section{Conclusions}

Ring diagram analysis yields the horizontal components of velocity in
the region $r>0.97R_\odot$. To test the validity of these results we 
compare the average longitudinal velocity with the rotation rate
inferred from inversion of global p-modes. Similarly, the inferred velocity
at the surface is compared with Doppler measurements.
It is found that the average longitudinal velocity agrees reasonably well
with the rotation rate inferred from inversion of global p-modes.
Similarly, the meridional component of velocity at the solar surface
agrees with that inferred by Doppler measurements.

The shear layer just below the solar surface is clearly seen in our results.
It appears
that this layer probably consists of two parts, the upper layer confined
to a depth of up to 4 Mm, where the gradient in rotation velocity is quite
steep and does not change sign up to a latitude of $60^\circ$.
The meridional component of velocity is found to be roughly independent of
depth in this shear layer. This shear layer roughly coincides with the
hydrogen ionization zone in solar models. Density increases by more than
two orders of magnitude in this layer.
There is some ambiguity in region very close to solar surface (depth $<2$ Mm)
as this region is not properly resolved in our study. It would be interesting
to use high resolution Dopplergrams to study flow velocities in this region
where superadiabatic gradient could be large. The second shear layer
below a depth of 4 Mm has smaller gradient in the rotation
component, and the gradient of rotation velocity appears to change sign
around a latitude of $55^\circ$, as is also found in rotation rate inferred
from global p-modes. The meridional components of velocity show some
variation in this lower shear layer. The second shear layer coincides with
the ionization zones of helium in solar models. There may be some distinction
between the layers covering the first and second ionization zones of helium,
but the difference
in velocity gradient between these two zones is not very clear at all latitudes.

The  velocity of the zonal flows  in the outermost region is similar to
that estimated from the splitting coefficient for the
global f-modes as well as the surface measurements (Hathaway et al.~1996).
The antisymmetric component of the rotation velocity is small
($<5$ m/s) and it is not clear if it is significant at all latitudes and
depths. Near the surface the antisymmetric component appears to be
significant for latitudes of 20--30$^\circ$.
In deeper layers, where the errors are larger, it is not clear if the zonal
flows or the antisymmetric component are significant.

The dominant signal in the meridional velocity is the meridional
flow which varies with latitude and has a maximum magnitude of about 35 m/s.
The dominant component of meridional flow has the form $\sin(2\theta)$,
but higher order components are also significant at intermediate depths.
In particular, the $P_4^1(\phi)$ component suggested by Durney~(1993) is
found to have an amplitude of about 3--4 m/s at intermediate depths.
The amplitude of the dominant component is about 30 m/s at the surface
which is in agreement with other measurements.
The north-south symmetric component of the meridional flow is generally
small and comparable to the error estimates,
except possibly in the outermost layers.
There is no change in sign of meridional velocity with depth up to 21 Mm.

\acknowledgments

This work  utilizes data from the Solar Oscillations
Investigation / Michelson Doppler Imager (SOI/MDI) on the Solar
and Heliospheric Observatory (SoHO).  SoHO is a project of
international cooperation between ESA and NASA.
The authors would like to thank the SOI Science Support Center
and the SOI Ring Diagrams Team for assistance in data
processing. The data-processing modules used were
developed by Luiz A. Discher de Sa and Rick Bogart, with
contributions from Irene Gonz\'alez Hern\'andez and Peter Giles.
SB is supported by an AMIAS fellowship.

\vfill\eject

\begin{figure}
\plotfiddle{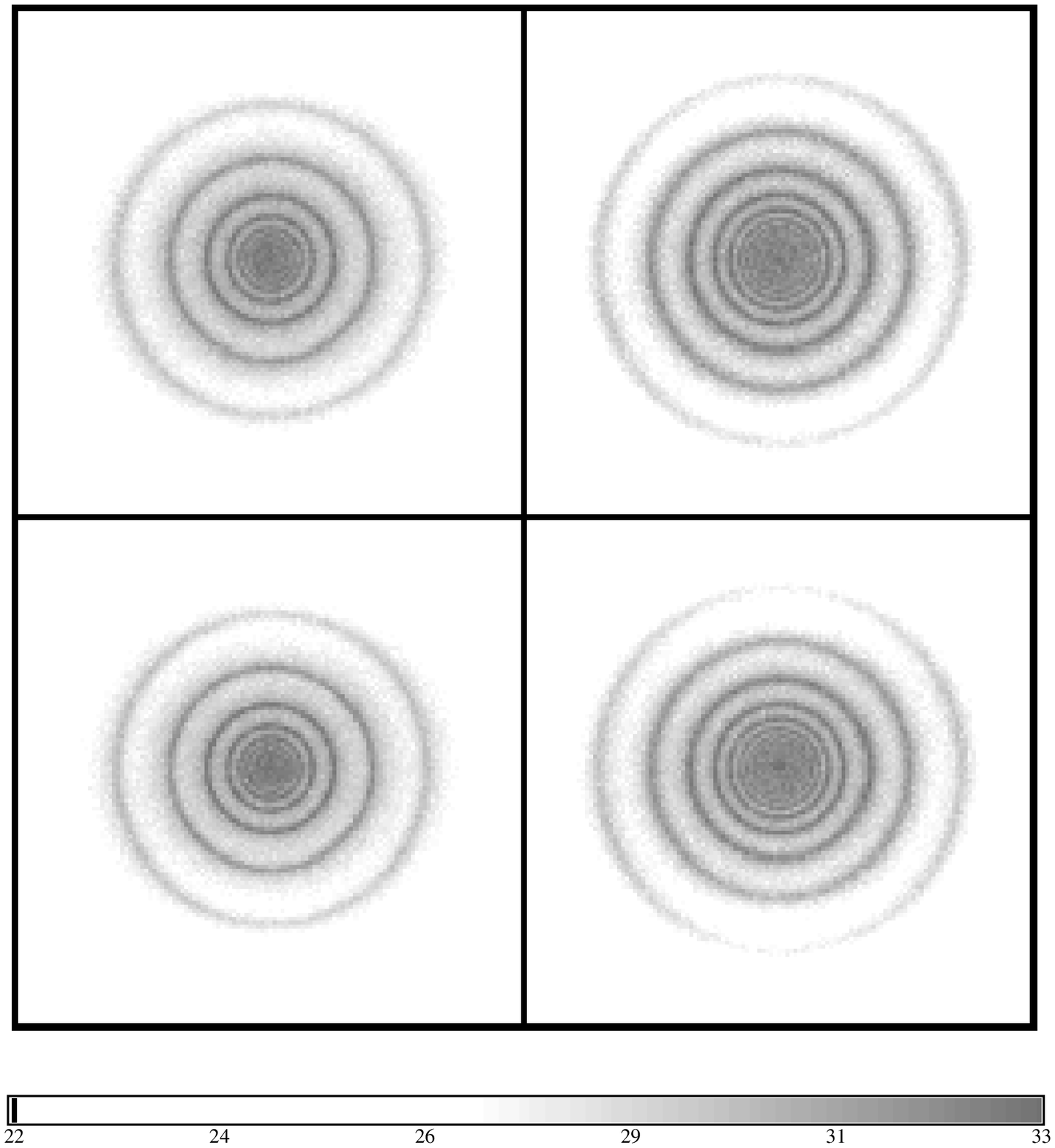}{9 true cm}{0}{70}{70}{-215}{-70}
\figcaption{Sample (logarithmic) power spectra
as a function of $k_x$ and $k_y$
at a fixed frequency. The top panels are for the average spectrum centered
at equator, at frequencies of
around 3 mHz (left) and 4 mHz (right). The bottom panels are for
the spectrum centered at $45^\circ$N latitude, at frequencies of around
3 mHz (left) and 4 mHz (right). The scale is marked with the logarithm
of the power.
}
\end{figure}
\vfill\eject

\begin{figure}
\plotone{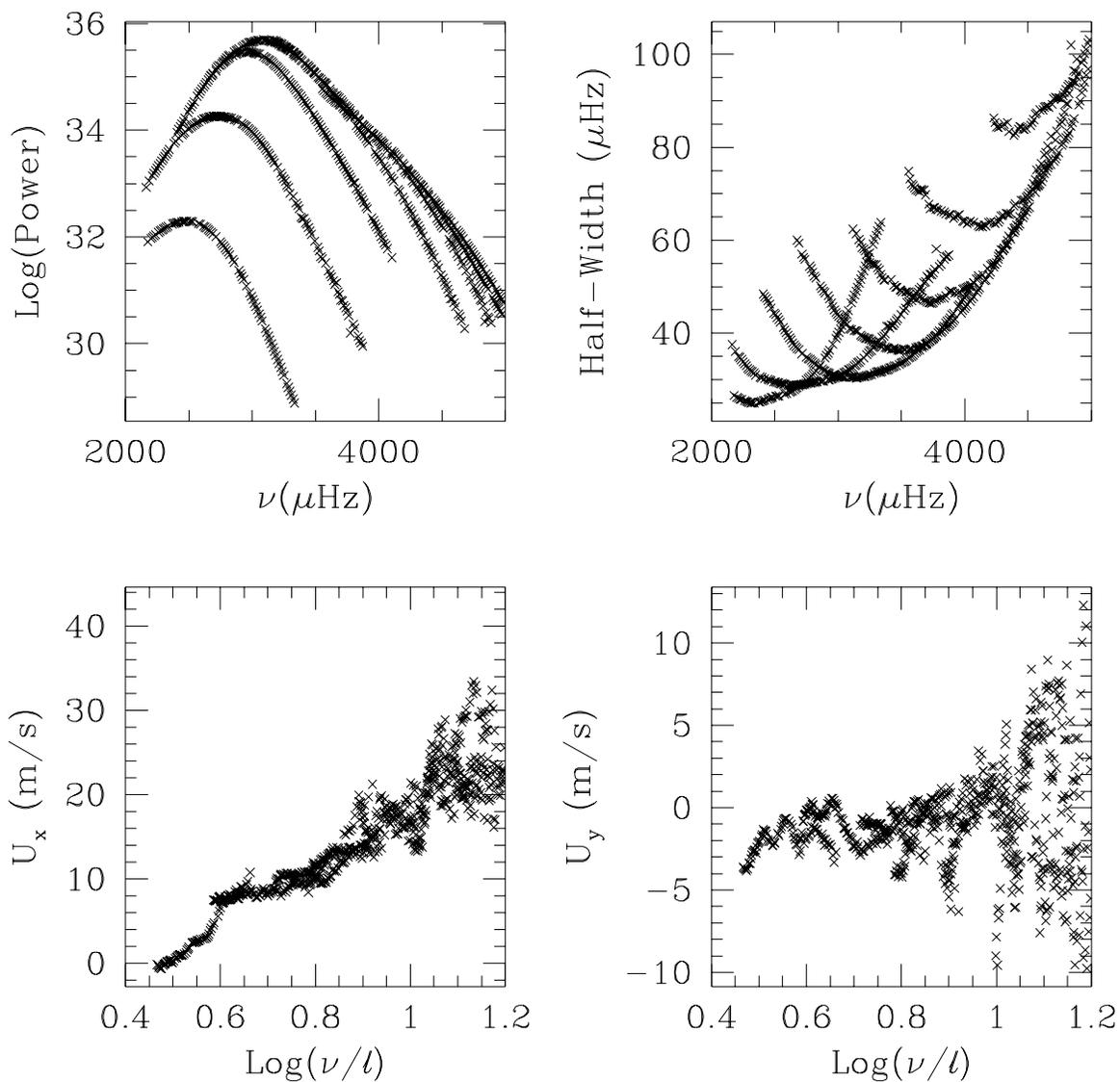}
\figcaption{ The fitted parameters for the summed spectra centered at
the equator. This figure shows the mean logarithmic power ($A_0-2\ln w$),
the half-width ($w_0$), and the average horizontal velocity $U_x, U_y$.}
\end{figure}
\vfill\eject

\begin{figure}
\plotfiddle{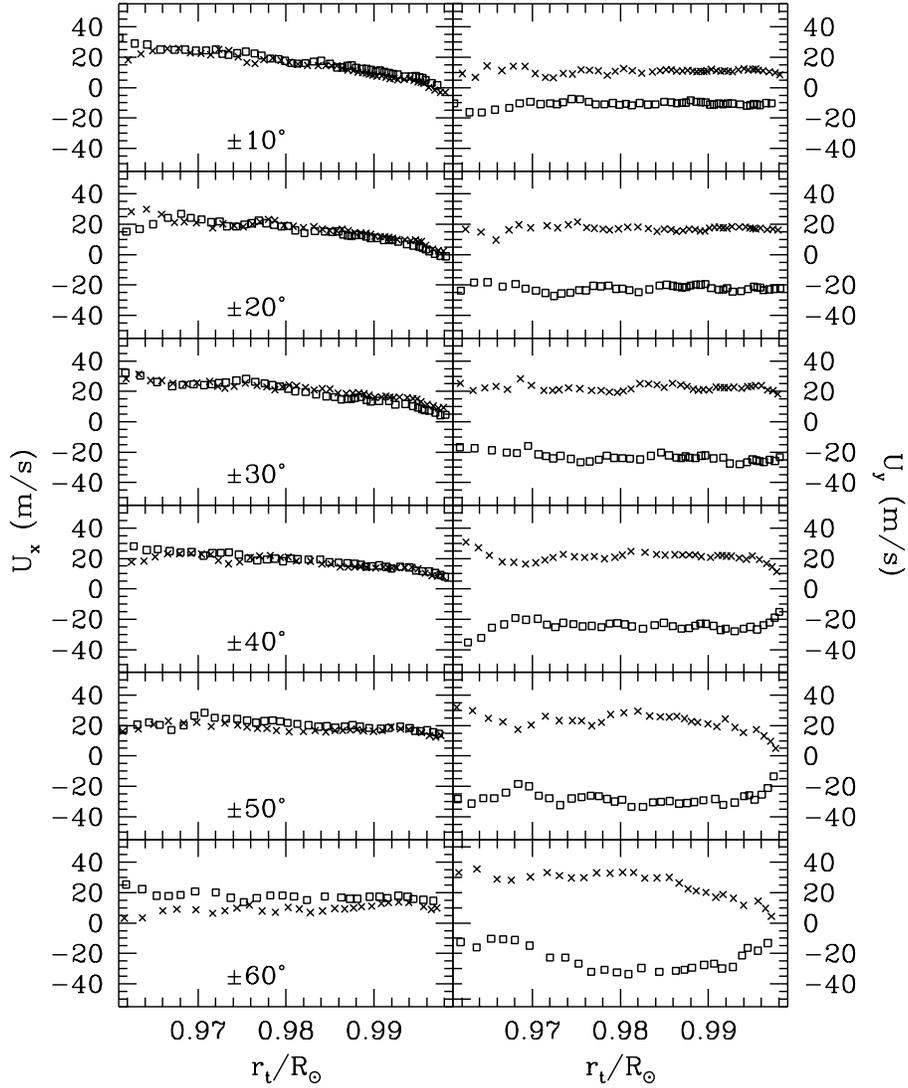}{10 true cm}{0}{80}{80}{-240}{-150}
\figcaption{The fitted velocity terms for the summed power spectra at
different latitudes  plotted as a function of the lower turning point
($r_t$) of the modes. In each panel the crosses mark the fitted
velocity for northern hemisphere while the open squares mark that for
the southern hemisphere. For clarity, the modes are averaged in groups of
15 and the error bars are not shown.
The latitudes are marked in the left panel.}
\end{figure}
\vfill\eject

\begin{figure}
\plotone{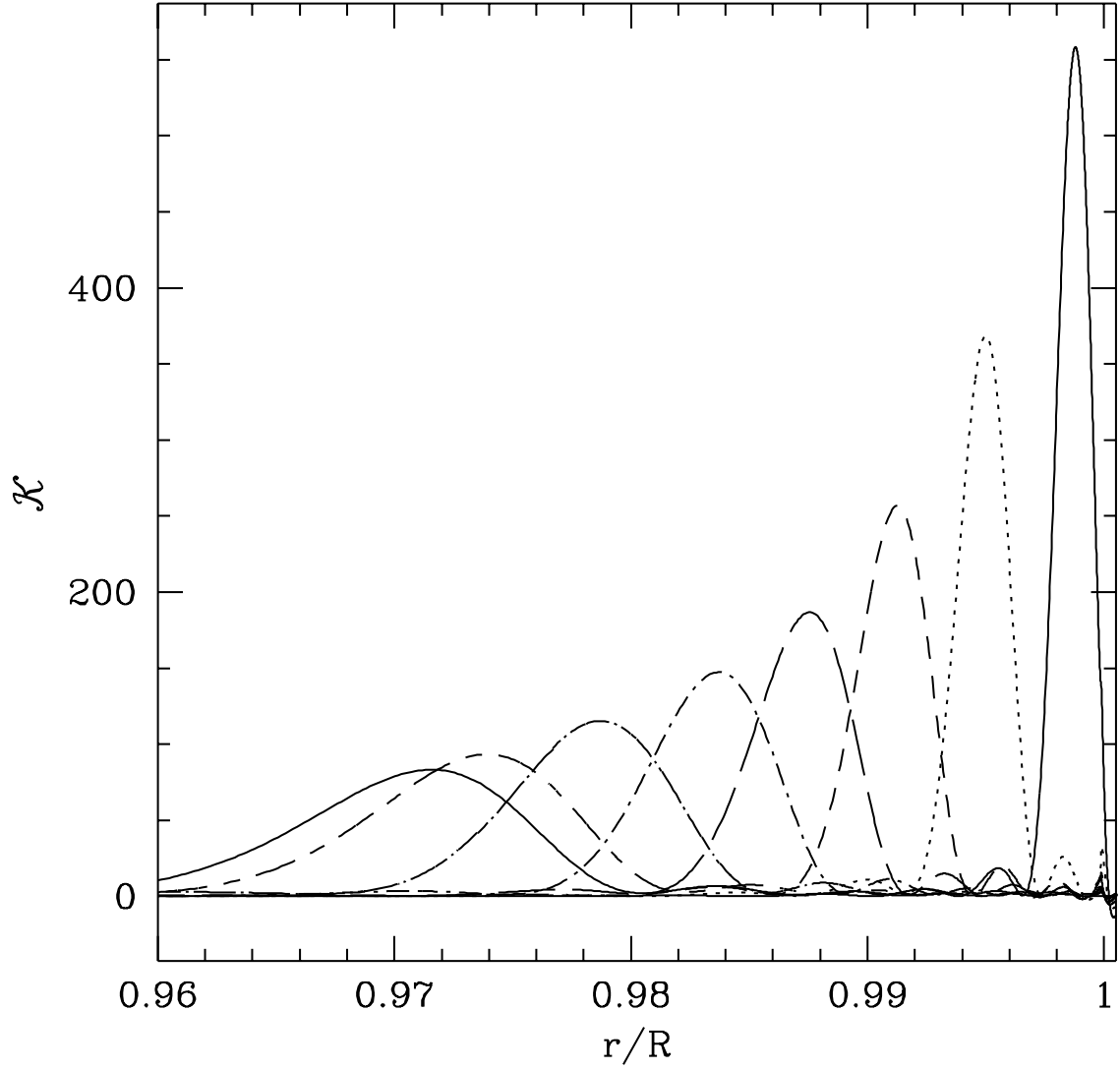}
\figcaption{A sample of the averaging kernels obtained for the OLA
inversions.  The averaging kernels shown are for inversions at
0.9705, 0.9731, 0.9782. 0.9833, 0.9872, 0.9910, 0.9949 and 0.9987
$R_\odot$ for $45^\circ$ N latitude.
}
\end{figure}
\vfill\eject

\begin{figure}
\plotone{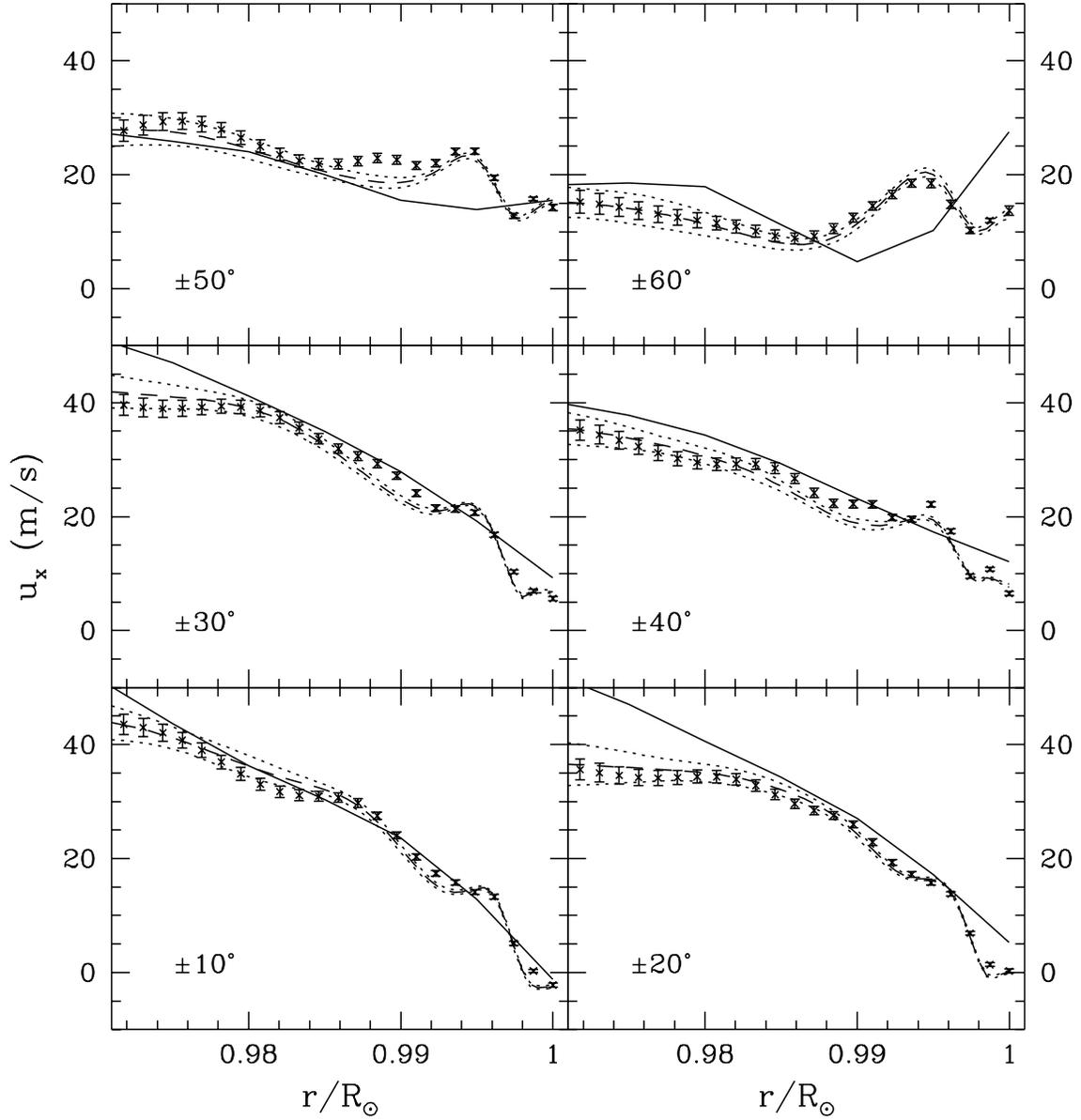}
\figcaption{The latitudinally symmetric part [$(u_N+u_S)/2$] of
the average horizontal velocity at
different latitudes (dashed line for RLS and crosses for OLA),
compared with the rotation velocity obtained
from inversion of splitting-coefficients plotted  after subtracting out the
surface rotation rate used in tracking each region
(continuous line).}
\end{figure}
\vfill\eject

\begin{figure}
\plotone{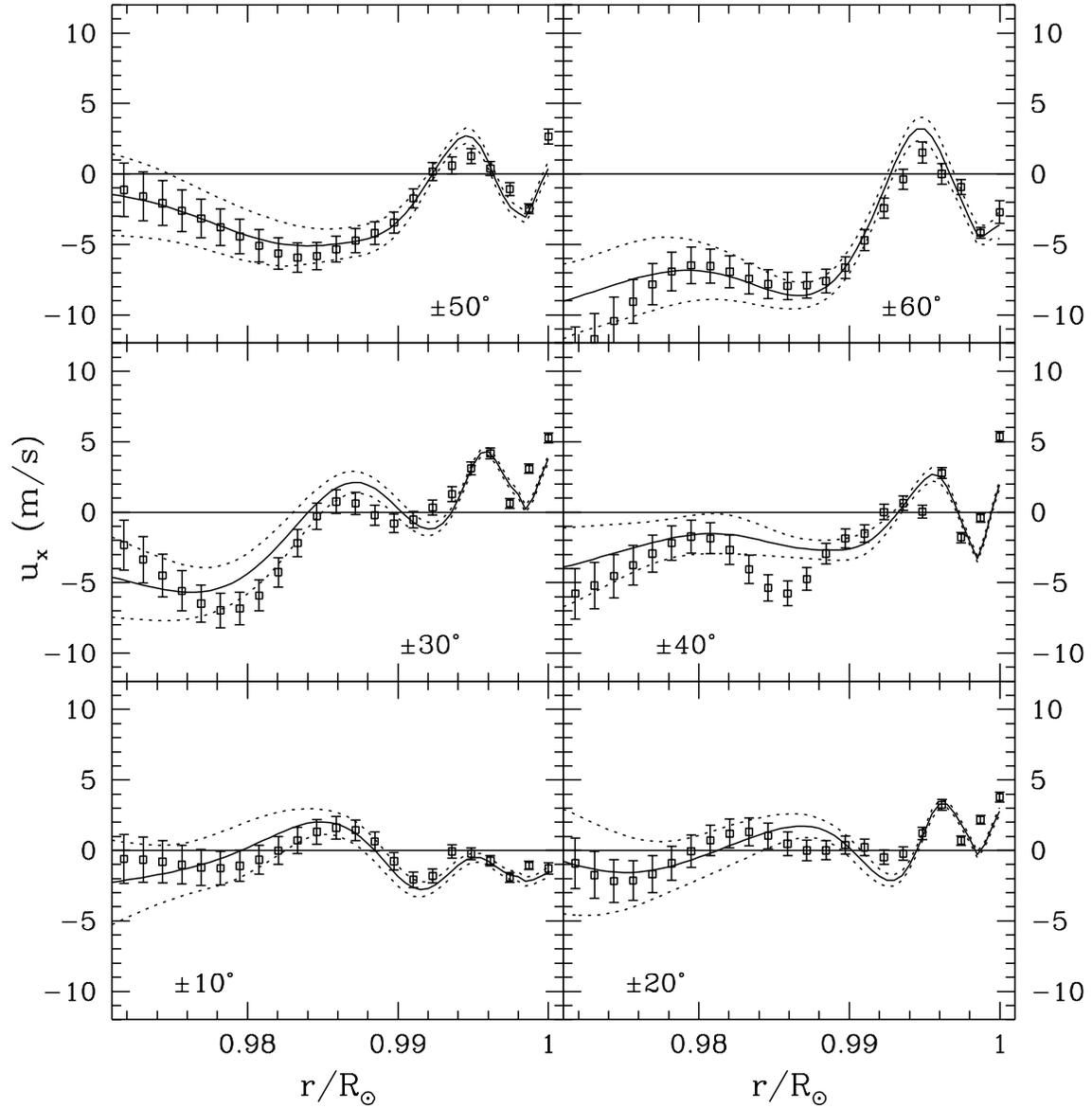}
\figcaption{The antisymmetric component [$(u_N-u_S)/2$] of the
rotation velocity plotted as a function of depth for
various latitudes. The continuous lines are RLS results with dotted
lines marking the $1\sigma$ error limits and open squares are OLA results.}
\end{figure}
\vfill\eject

\begin{figure}
\plotone{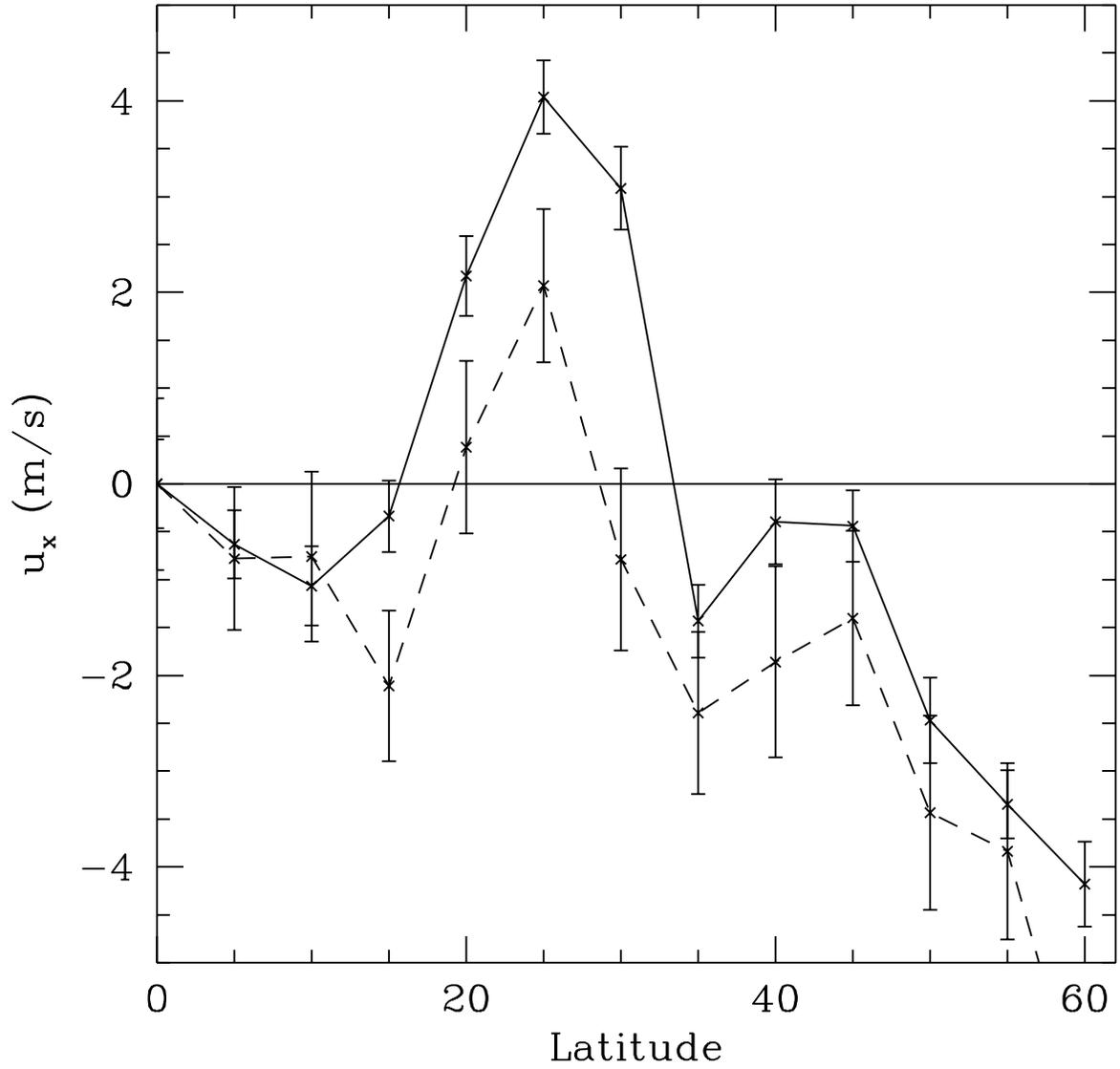}
\figcaption{The antisymmetric component [$(u_N-u_S)/2$] of the
rotation velocity plotted as a function of latitude for
$r=0.997R_\odot$ (continuous line) and $r=0.990R_\odot$ (dashed line).
These results are obtained using OLA technique.}
\end{figure}
\vfill\eject

\begin{figure}
\plotfiddle{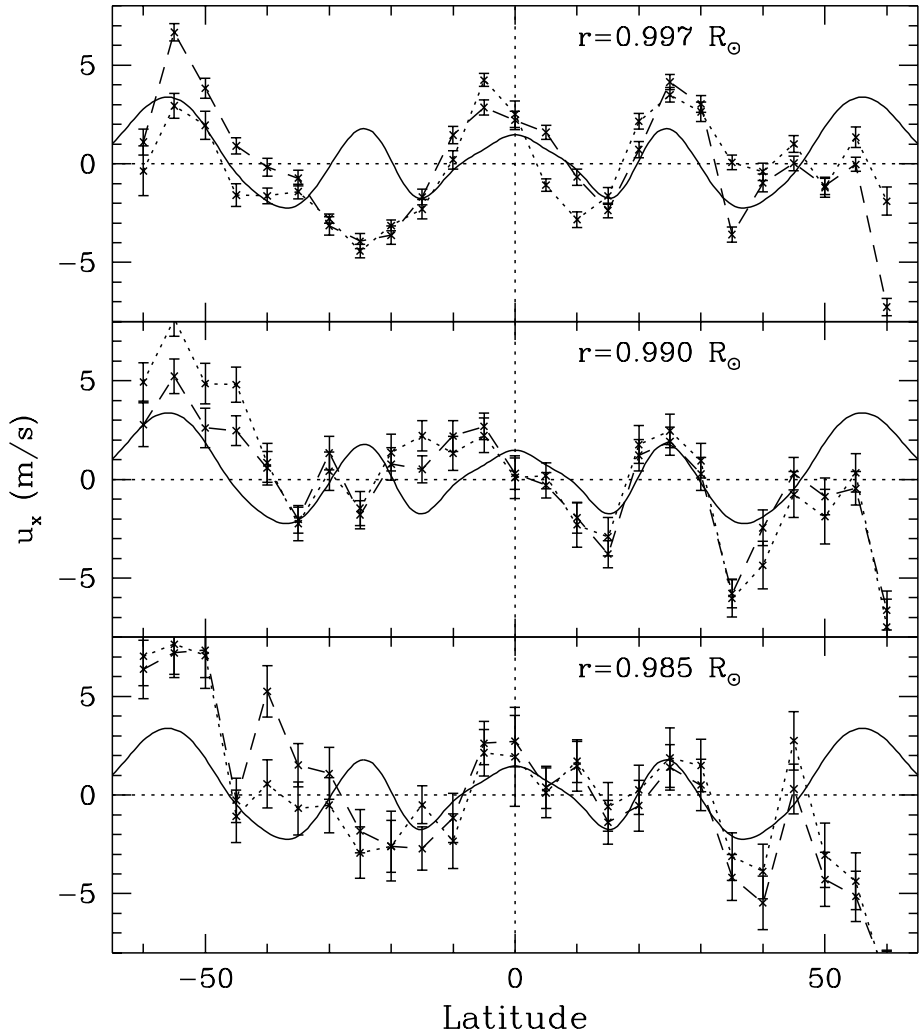}{9 true cm}{0}{85}{85}{-290}{-190}
\figcaption{The zonal flow, i.e., the residual  rotation velocity left
after removing a smooth component (obtained by a three term fit)
at a few depths using OLA (dashed lines)
and RLS (dotted lines) inversions. The depths are marked in the panels. 
The continuous lines represent the average zonal flow velocity as inferred
from  f-modes  using the 360 day MDI splitting coefficients. The dotted 
horizontal and vertical lines mark the two axes.}
\end{figure}

\vfill\eject

\begin{figure}
\plotfiddle{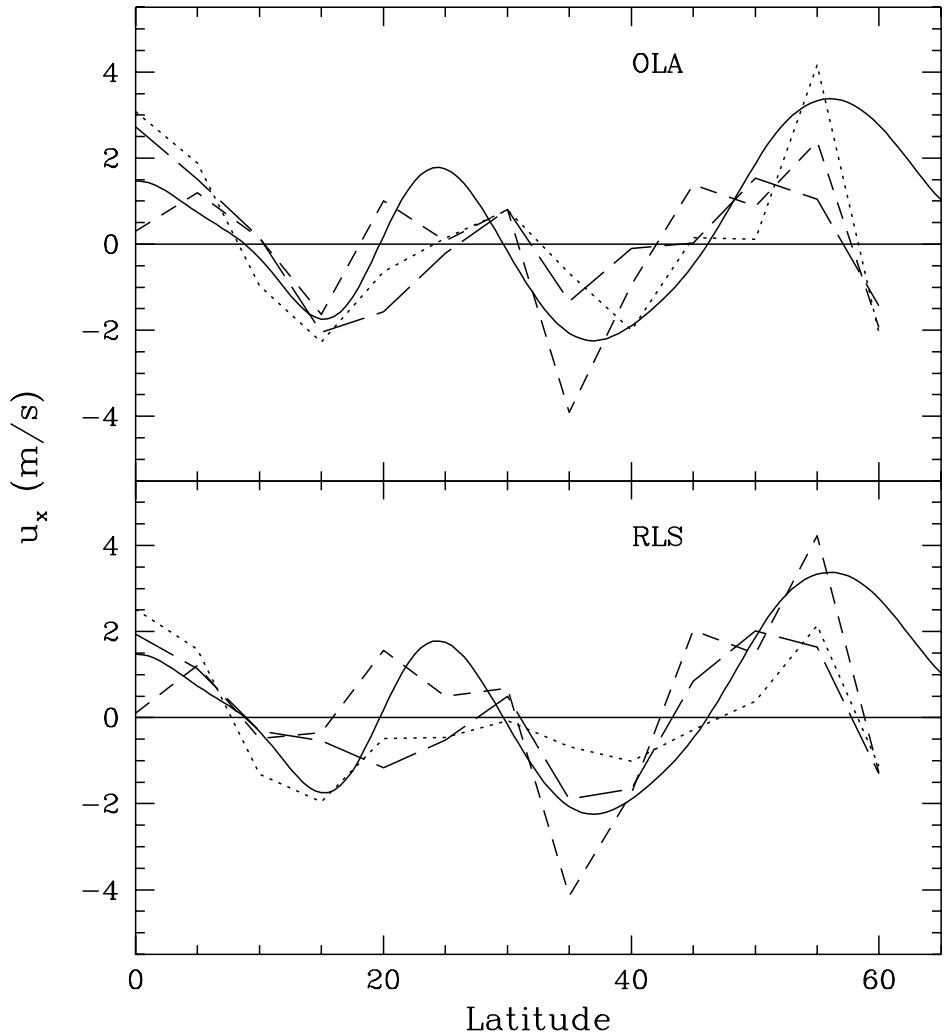}{9 true cm}{0}{85}{85}{-285}{-190}
\figcaption{The zonal flow
at a few depths obtained by  OLA (upper panel)
and RLS (lower panel) inversions. This figure shows the 
north-south symmetric component of the zonal flow shown in Fig.~8.
The dotted lines are for $r=0.997R_\odot$, the short dashed lines
are for $r=0.990R_\odot$ and the long dashed lines are for $r=0.985R_\odot$.
Error bars are not shown for clarity.
The continuous lines represent the average zonal flow velocity as inferred
from the f-modes using the 360 day MDI splitting coefficients.}
\end{figure}
\vfill\eject

\begin{figure}
\plotfiddle{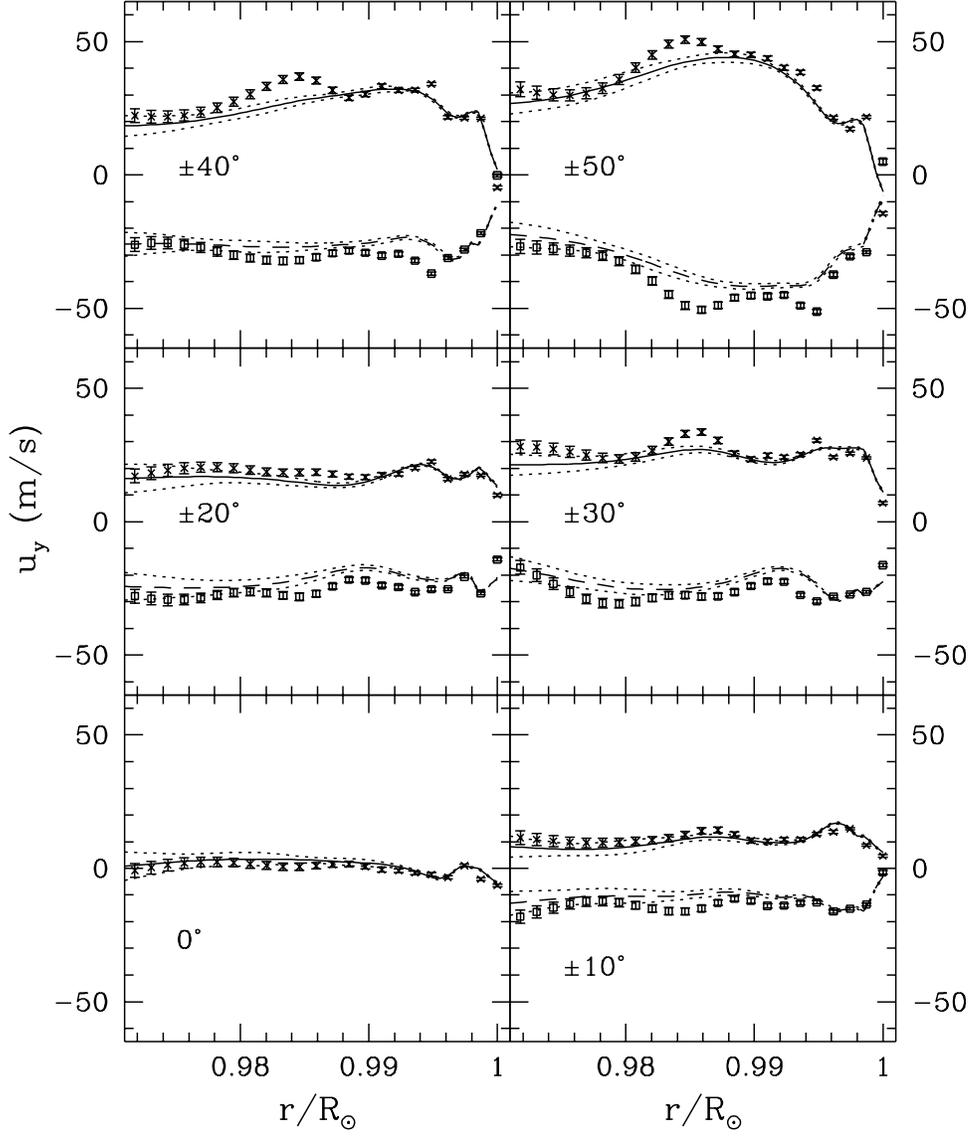}{10 true cm}{0}{83}{83}{-255}{-150}
\figcaption{Meridional velocity at different latitudes
plotted as a function of depth.
The results of RLS inversions are shown by continuous lines for
northern latitudes and dashed lines for  southern latitudes,
with the dotted lines showing the
$1\sigma$ error limits.  The crosses for northern latitudes
and squares for
southern latitudes, mark the results of OLA inversions. 
}
\end{figure}
\vfill\eject

\begin{figure}
\plotone{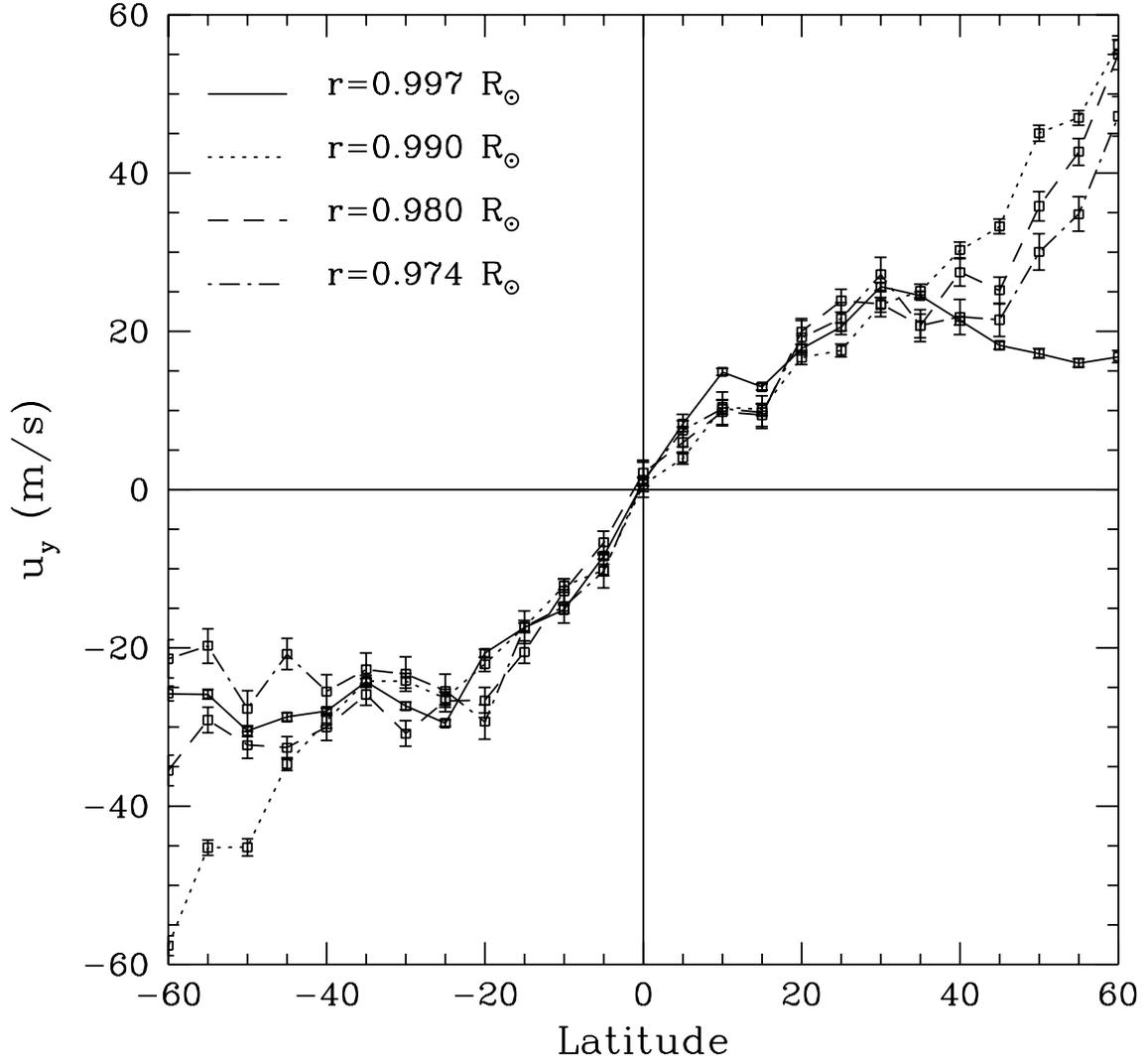}
\figcaption{Meridional velocity at different depths
plotted as a function of latitude. These results have been obtained
using the  OLA technique for inversion.
}
\end{figure}
\vfill\eject

\begin{figure}
\plotone{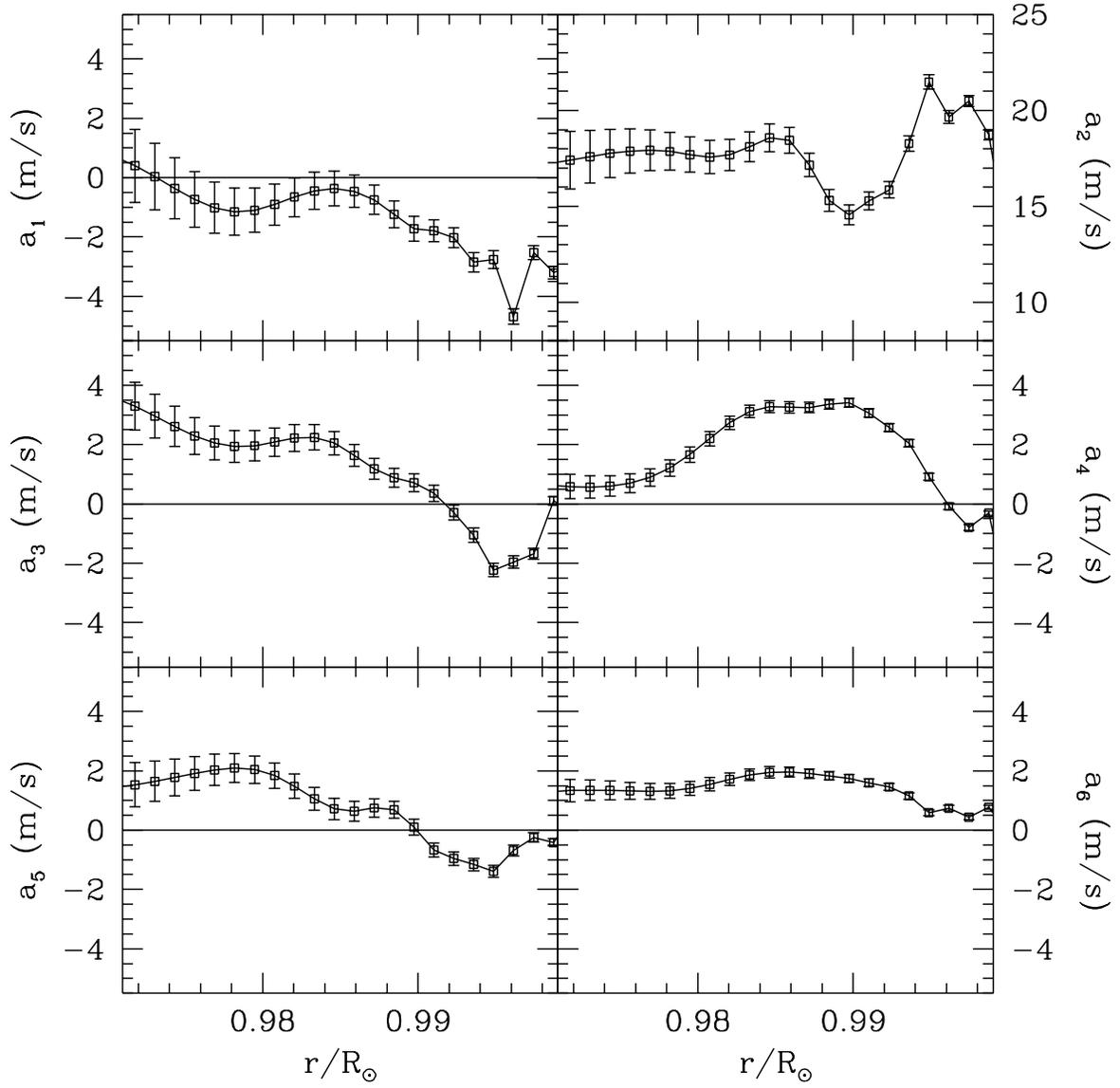}
\figcaption{Amplitude of various components of meridional velocity as a
function of depth. The results are obtained using OLA technique.
}
\end{figure}

\end{document}